\documentclass[preprint,showpacs,preprintnumbers]{revtex4}
\usepackage{graphicx}
\usepackage{dcolumn}
\usepackage{bm}

\begin{document}

\title{Negative dimensional approach to evaluating real integrals}
\author{Alfredo Takashi Suzuki}
\affiliation{Instituto de F\'{\i}sica Te\'orica,\\
Universidade Estadual Paulista, \\
Rua Pamplona, 145 \\
01405-900 Sao Paulo, SP - Brazil}
\date{\today}

\begin{abstract}
In solving the differential equation for a non damped harmonic oscillator
one meets, after subjecting the equation to a Fourier transformation, an
integration in the complex $\omega$ plane. In most cases such an integral is
evaluated by calculating residues together with some physical input such as
the principle of causality to define which pole residues are relevant to the
physical problem. For this kind of application, Cauchy's theorem or residue
theorem can be applied to evaluate certain real integrals. Here we present
an alternative approach based on the concept of negative dimensional
integration to treat such integrals and give an specific example on how this
is accomplished.

\vspace{5cm}
\end{abstract}

\pacs{02.10.Ud; 02.30.-f; 02.30.Cj}
\maketitle


Mathematics Subject Classification (2000): 28E05 Nonstandard Measure Theory







\section{Introduction}

In a textbook on complex variables we may find real definite integrals of
the type 
\begin{equation}  \label{I0}
I_0=\int_{-\infty}^{+\infty} \! \frac{P(x)}{Q(x)}\,dx,
\end{equation}
where $P(x)$ and $Q(x)$ are polynomials in $x$ satisfying the following two
requisites: (a) the degree of $Q(x)$ must exceed the degree of $P(x)$ by at
least 2, to ensure convergence and vanishing of the integral over the
semicircle $C_{\mathrm{R}}$ in the limit of $\mathrm{R}$ extended to
infinity; and (b) $Q(x)$ is a polynomial with no real zeros. This second
requirement can be somehow relaxed by special modifications in order to make
the integral become defined in such a case. The standard result for such
integrals is 
\begin{equation}  \label{I0_a}
I_0=\int_{-\infty}^{+\infty} \! \frac{P(x)}{Q(x)}\,dx = 2\pi\,i\sum_{+}%
\mathrm{Res}
\end{equation}
where $\sum_{+}\mathrm{Res}$ is the sum of the residues of the integrand in
the complex upper half-plane.

This can be applied straightforwardly to the integral 
\begin{equation}
I_{\mathrm{cos}} = \int_{-\infty}^{+\infty} \frac {\cos x}{x^2+a^2}\,dx = 
\frac{\pi}{a}\,\mathrm{e}^{-a}.
\end{equation}
as the real part of the more general integral 
\begin{equation}  \label{I_exp}
I_{\mathrm{exp}} = \int_{-\infty}^{+\infty} \frac {\mathrm{e^{ix}}}{x^2+a^2}%
\,dx = \frac{\pi}{a}\,\mathrm{e}^{-a}.
\end{equation}

Now, it is a well-known result that all the positive dimensional polynomial
integrals of the form below vanishes: 
\begin{equation}
I_{\mathrm{pol}} = \int_{-\infty}^{+\infty}(x^2)^n\,dx = 0, \qquad n\geq 0.
\end{equation}

However, when we allow an analytic continuation to negative dimensions, the
above integral has a non vanishing and nontrivial result \cite{ricotta}: 
\begin{equation}
I_{\mathrm{pol}} = \int_{-\infty}^{+\infty}(x^2)^n\,d^\star x = (-1)^n\, n!\,%
\sqrt{\pi}\,\delta_{n+\frac{1}{2},\,0}.
\end{equation}
where we have written the measure $d^\star x$ to remind ourselves that we
are in negative dimensional space.

In the next section we apply this concept of negative dimensional space to
evaluate (\ref{I_exp}).

\section{Negative Dimension Integration Method (NDIM)}

Since the integrand in (\ref{I_exp}) has an exponential function which is
not in the form a polynomial, first of all, we need to convert it in a
polynomial form to apply the negative dimensional technique. We can do this
by using the series expansion for the exponential function, namely, 
\begin{equation}  \label{I_G}
I_{\mathrm{exp}} = \sum_{m=0}^{+\infty} \frac{i^m}{m!}\int_{-\infty}^{+%
\infty}\,dx\,\frac{x^m}{x^2+a^2}
\end{equation}

Let us now define the Gaussian generating functional of the negative
dimensional integration \cite{labtesting} corresponding to the integral in (%
\ref{I_G}), namely, 
\begin{equation}  \label{generating}
\mathcal{G}=\int_{-\infty}^{+\infty}\!dx\;\mathrm{{e}^{-\alpha x}\,{e}%
^{-\beta(x^2+a^2)}}
\end{equation}
which following the NDIM procedure we write as 
\begin{equation}  \label{sum1}
\mathcal{G} = \sum_{r,s=0}^{+\infty} (-1)^{r+s}\frac{\alpha^r \beta^s}{r!\,s!%
}\,\int_{-\infty}^{+\infty}\,d^\star x\,x^{r}(x^2+a^2)^s
\end{equation}

Observe now that for $s=-1$ the integral 
\begin{equation}
I_0^{\star}(s) \equiv \int_{-\infty}^{+\infty}\,d^\star x\,x^{r}(x^2+a^2)^s
\end{equation}
is exactly the integral that appears in (\ref{I_G}).

This means that after $I_0^{\star}(s)$ is performed with positive values of $%
s$ and negative dimension, we need to analytic continue it to negative
values of $s$ and positive dimension.

On the other hand, the Gaussian generating functional integral (\ref%
{generating}) can be evaluated, yielding 
\begin{equation}
\mathcal{G} = \sqrt {\frac{\pi}{\beta}}\,\mathrm{e}^{-\beta a^2+\frac {%
\alpha^2}{4\beta}}
\end{equation}

Expanding the exponential function $\mathrm{e}^{-\beta a^2+\frac {\alpha^2}{%
4\beta}}$ in power series, we get 
\begin{equation}  \label{sum2}
\mathcal{G} = \pi^{1/2} \sum_{k,l=0}^{+\infty}\frac{(-1)^k(a^2)^k}{%
4^l\;k!\,l!}\,\alpha^{2l}\, \beta^{k-l-1/2}
\end{equation}

Now, comparing (\ref{sum1}) and (\ref{sum2}) term by term, we have to have
the following constraints satisfied: 
\begin{eqnarray}
2l & = & r  \nonumber \\
k-l-\frac {1}{2} & = & s,
\end{eqnarray}

Expressing the different factorials that appear in terms of gamma functions,
i.e., $p! = \Gamma(p+1)$ and using the Pochhammer's symbols for ratios of
gamma functions, i.e. 
\begin{equation}
(p)_q \equiv \frac{\Gamma(p+q)}{\Gamma(p)}
\end{equation}
we get 
\begin{equation}
I_0^{\star}(s) = \frac{(-\pi)^{1/2}(a)^{r+2s+1}}{2^r\;(r+1)_{-r/2}%
\,(s+1)_{r/2+1/2}}
\end{equation}

We need now to analytic continue this result to allow for negative values of 
$s$ and positive dimension. Note here that only the exponent $s$ needs to be
analytic continued to allow for negative values, since $r$ is always
positive \cite{tensorial}. This is accomplished by using the property of the
Pochhammer's symbols, namely, 
\begin{equation}
(p)_q = \frac{(-1)^{-q}}{(1-p)_{-q}}
\end{equation}
for the $s$ bearing Pochhammer, so that 
\begin{equation}
I_0^{\star, \mathrm{AC}}(s) = \frac{i^r}{2^r}\;\pi^{1/2}\;(a)^{r+2s+1}\frac{%
(-s)_{-r/2-1/2}}{(r+1)_{-r/2}}
\end{equation}

This, for the particular value of $s=-1$ which is of interest for us, yields 
\begin{equation}
I_0^{\star, \mathrm{AC}}(-1) = \frac{\pi}{a}(-a)^r
\end{equation}
which introduced into (\ref{I_G}) gives 
\begin{equation}
I_0 = \sum_{m=0}^{+\infty} \frac{\pi}{a}\frac{(-a)^m}{m!} = \frac {\pi}{a}\;%
\mathrm{e}^{-a}
\end{equation}
which agrees with the one obtained through the use of complex plane residue
theorem.

\section{Conclusions}

In this work we have considered a class of real integrals which usually is
performed with the help of Cauchy's residue theorem. We have presented here
an alternative method whereby this type of real integrals can also be
performed without difficulty.

\end{document}